\documentclass[a4, 14pt]{article}
\usepackage{amssymb,epic,eepic} %
\usepackage{color,psfig,epic,eepic,epsf,epsfig,amsfonts,amssymb,amsmath,latexsym}
\usepackage[latin1]{inputenc}
\author{ Igor Bjelakovi\'c$^{1}$  \and Jean-Dominique Deuschel$^{1}$
  \and Tyll Kr\"uger$^{1,2}$  \and Ruedi Seiler$^{1}$ \and
  Rainer Siegmund-Schultze$^{1,3}$  \and
  Arleta Szko\l a$^{1,4}$\footnote{e-mail:\{igor, deuschel, tkrueger,
    seiler, siegmund,
    szkola\}@math.tu-berlin.de}   \\
$^{1}${\footnotesize Technische Universit\"at Berlin} \\
{\footnotesize Fakult\"at II - Mathematik und Naturwissenschaften} \\
{\footnotesize Institut f\"ur Mathematik MA 7-2} \\
{\footnotesize Stra\ss e des 17. Juni 136,} 
{\footnotesize 10623 Berlin, Germany}\\ \\
$^{2}${\footnotesize Universit\"at Bielefeld}\\
{\footnotesize  Fakult\"at f\"ur Mathematik}\\
{\footnotesize Universit\"atsstr. 25, 33619 Bielefeld, Germany}\\ \\
$^{3}${\footnotesize Technische Universit\"at Ilmenau}\\
{\footnotesize Institut f\"ur Mathematik}\\
{\footnotesize 98684 Ilmenau, Germany}\\ \\
$^{4}${\footnotesize Max Planck Institute for
Mathematics in the Sciences}\\
{\footnotesize Inselstrasse 22, 04103 Leipzig, Germany}
}

\title{A Quantum Version of Sanov's Theorem}

\newcommand{\cc}{{\mathbb C}}

\newcommand{\nn}{{\mathbb N}}
\newcommand{\idn}{\mathbf{1}}

\newcommand{\eps}{{\varepsilon}}        
\newcommand{\vphi}{{\varphi}}           

\newtheorem{theorem}{Theorem}[section]         
\newtheorem{lemma}[theorem]{Lemma}             
\newtheorem{proposition}[theorem]{Proposition} 

\begin{document}

\maketitle

\begin{abstract}
We present a quantum extension of a version of
Sanov's theorem focussing on a hypothesis testing aspect of the theorem:
There exists a sequence of typical subspaces for a given set $\Psi$ of stationary quantum product states asymptotically separating them from another fixed stationary product state. Analogously to the classical case, the exponential separating rate is equal to the infimum of the quantum relative entropy with respect to the quantum reference state over the set $\Psi$. However, while in the classical case the separating subsets can be chosen universal, in the sense that they depend only on the chosen set of i.i.d. processes, in the quantum case the choice of the separating subspaces depends additionally on the reference state. 
\end{abstract}

\section{Introduction}

In this article we present a natural quantum version of the classical Sanov's
theorem as part of our attempt to explore basic concepts and results 
at the interface of classical information theory and stochastics from the
point of view of quantum information theory. 

Among those classical results a crucial role plays the 
\emph{Shannon-McMillan-Breiman theorem} (SMB theorem) which clarifies the
concept of \emph{typical subsets}, yielding the rigorous background for
asymptotically optimal lossless data compression. It says 
that a long $n$-block message string from an 
ergodic data source belongs most likely to a typical
subset (of the generally very much larger set of all possible messages). The
cardinality within the sequence of these typical sets grows with the length $%
n$ of the message at an exponential rate given by the \emph{Shannon entropy
rate} of the data source. As a general rule, when passing to the quantum
situation, the notion of a typical subset has to be replaced by that of a 
\emph{typical subspace} (of an entire Hilbert space describing the pure $n$%
-block states of the quantum data source), with the dimension of that
subspace being the quantity growing exponentially fast at a rate given now
by the \emph{von} \emph{Neumann entropy rate }as $n$ goes to infinity (cf.
\cite {jozsa} in the i.i.d. situation, \cite {wir1}, \cite{wir2} in the general ergodic
case).

We know from the classical situation that typical subsets have even more
striking properties, when chosen in the right way: For a given alphabet $A$
and a given entropy rate there is (a bit surprisingly) a \emph{universal }%
sequence of typical subsets growing at the given rate \emph{for all ergodic
sources} which do not top the given entropy rate (this result has been
generalized to the quantum context by Kaltchenko and Yang \cite{ky}).
Moreover, for any ergodic data source $P$ we can find a sequence of typical
subsets growing at the rate given by the entropy and at the same time \emph{%
separating }it exponentially well from any i.i.d. (reference) data source
$Q$ in
the sense that the $Q$-probability of the entire $P$-typical subset goes to
zero at an exponential rate given by the \emph{relative entropy rate} $%
h(P,Q).$ Furthermore, the relative entropy is the best achievable (optimal)
separation rate. This assertion which gives an operational interpretation of
the relative entropy is Stein's lemma. We mention that the i.i.d. condition
concerning the \emph{reference} source cannot be weakened too much, since there are
examples where even the relative entropy has no asymptotic rate, though the
reference source is 
very well mixing (B-process, cf. \cite{shields2} ). A quantum
generalization of this result can be found in \cite{ogawa} for the case that both
sources are i.i.d., and in \cite{bs1} for the case of a general ergodic quantum
information source. This result was mainly inspired by \cite{petz}, where complete
ergodicity was assumed and optimality was still left open.


From the viewpoint of information theory or statistical hypothesis testing
the essential assertion of \emph{Sanov's theorem} is that it represents a
universal version of Stein's lemma by saying that for a set $\Omega$ of
i.i.d.  sources there exists a \emph{common choice} of the typical set such
that the probability with respect to  the i.i.d.  reference source $Q$ goes
to zero at a rate given by $\inf_{P\in \Omega }h(P,Q).$  

Originally Sanov's theorem is of course a result on large deviations of
\emph{empirical distributions} (cf. \cite{Sanov}, \cite{Deuschel}). It is the information-theoretical
viewpoint taken here which suggests to look at it as a large deviation
principle for typical subsets. With the main topic of this paper being a
quantum theorem of \ Sanov type, it is especially appealing to shift the
focus from empirical distributions to typical subspaces, since the notion of an
\emph{individual }quantum message string is at least problematic, and as
will be seen by an example, a reasonable attempt to define something like a
quantum empirical distributions via partial traces leads to a separation
rate worse than the relative entropy rate (see the last section).

Another aspect of the classical Sanov result has to be modified for the
quantum situation: The typical subspace will no longer be universal for all
i.i.d. reference sources, but has to be chosen in dependence of the
reference source. So only 'one half' of universality is maintained when
passing to quantum sources, namely that which refers to the set $\Omega$%
. This will be demonstrated by an example in the last section.
The basic mechanism behinde this {\it{no go
result}} is - heuristically speaking: In the quantum setting even pure states
cannot be distinguished with certainty, while classical letters can. In our forthcoming paper \cite{wirnext} we extend the results given here to the case where only stationarity is assumed for the states in $\Omega$.

\section{A quantum version of Sanov's theorem}

Let $A$ be a finite set with cardinality $\# A=d$. By ${\cal P}(A)$ we
denote the set of probability distributions on $A$. The relative entropy
$H(P,Q)$ of a probability distribution $P \in {\cal P}(A)$ with respect to a distribution $Q$ is defined
as usual:

\begin{eqnarray}\label{rel-entr}
H(P,Q):=\left\{
  \begin{array}{ll}
\sum_{a \in A}P(a)(\log P(a) - \log Q(a)),&
 \textrm{if } 
P \ll Q
\\ 
\infty,&  \textrm{otherwise,}
  \end{array}
\right. 
\end{eqnarray} 
where $\log$ denotes the base $2$ logarithm. For the base $e$ logarithm we use the notation $\ln$.
The function $H(\cdot, Q)$ is continuous on ${\cal P}(A)$, if the reference
distribution $Q$ has full support $A$. Otherwise it is lower semi-continuous.
The relative entropy distance from the reference distribution $Q$ to a
subset $\Omega \in {\cal P}(A)$ is given by:
\begin{eqnarray}\label{rel-entr-dist}
H(\Omega, Q):= \inf_{P \in \Omega}H(P,Q).
\end{eqnarray}
Our starting point is the classical Sanov's theorem formulated 
from the point of view of hypothesis testing:
\begin{theorem}[Sanov's Theorem]\label{class-result} 
Let $Q \in {\cal
P}(A)$ and $\Omega \subseteq {\cal P}(A)$. There exists a sequence
$\{M_n\}_{n \in \nn}$ of subsets $M_n \subseteq A^n$ with 
\begin{eqnarray}
\lim_{n \to \infty}P^n(M_n)=1,\qquad \forall P \in \Omega,
\label{p-typical_set} 
\end{eqnarray} 
such that 
\begin{eqnarray} \lim_{n \to
\infty}\frac{1}{n} \log Q^n (M_n)= - H(\Omega,
Q).\label{classical_separation}  
\end{eqnarray}
Moreover, for each sequence of sets $\{\tilde M_{n}\}$ fulfilling
(\ref{p-typical_set}) we have 
\begin{equation*}
\liminf_{n\rightarrow \infty }\frac{1}{n}\log Q^{n}(\tilde M_{n})\geq -H(\Omega,Q),
\end{equation*}
such that $H(\Omega,Q)$ is the best achievable separation rate.
\end{theorem}

We emphasize that in the above formulation we omitted the assertion that the
sets $M_n$ can be chosen independently from the reference distribution $Q$.
However, as will be shown in the last section, in the quantum case this universality
feature is not valid any longer and Theorem \ref{class-result} is the
strongest version that has a quantum analogue.
It is an immediate consequence of Lemma (\ref{class-result2}) and
is related to the usual formulation of Sanov's theorem (\cite{Sanov},
see also Theorem 3.2.21 in \cite{Deuschel}) in terms of
empirical measures $P_{x^n}:=\frac{1}{n}\sum_{i=1}^{n}\delta_{x_{i}}$
for sequences $x^{n}:=\{x_{1},\ldots,x_{n}\}$ as follows:  By the strong law of large numbers, the sequence of empirical
distributions $\{P_{x^{n}}\}$ formed along an i.i.d. sequence $%
\{x_{1},x_{2},...\}$ of letters distributed according to a probability
measure $P\in \mathcal{P}(A)$ tends to $P$ almost surely. Hence for any
neighbourhood $U$ of $P$ we have the limit relation $\lim_{n\rightarrow
\infty }P^{n}(\{x^{n}:P_{x^{n}}\in U\})=1$ meaning that the sequence of sets 
$\{x^{n}:P_{x^{n}}\in U\}$ is typical for $P^{n}$. If $\Omega $ is an open
set, then we may choose $U$ as $\Omega $ and $\{x^{n}:P_{x^{n}}\in U\}$ is
universally typical for all $P^{n},P\in \Omega $. Now Sanov's theorem in its
traditional form says that $-\frac{1}{n}\log Q^{n}(\{x^{n}:P_{x^{n}}\in
\Omega \})\rightarrow H(\Omega ,Q)$. So it says that the (explicitely
specified) typical sets separate $Q^{n}$ exponentially fast from all $%
P^{n},P\in \Omega $ with the given order $H(\Omega ,Q)$. 

Passing to the quantum setting we substitute the set $A$ by a $C^*$-algebra
${\cal A}$ with dimension $\dim {\cal A}=d < \infty$ and the cartesian product $A^n:= \prod_{i=1}^{n}A$ by the tensor product ${\cal A}^{(n)}:= \bigotimes_{i=1}^n {\cal A} $.  We denote by ${\cal
S}({\cal A})$ the set of quantum states on the algebra of observables
${\cal A}$, i.e. $ {\cal S}({\cal A})$ is the set of positive functionals
$\vphi$ on ${\cal A}$ fullfilling the normalisation condition $\vphi(\idn)=1$. For $\vphi \in {\cal S}({\cal A})$ we mean by $\vphi^{\otimes n}$ a product state on ${\cal A}^{(n)}$.
The quantum relative entropy $S(\psi, \vphi)$ of the state $\psi \in {\cal
S}({\cal A})$ with respect to the reference state $\vphi \in {\cal S}({\cal
A})$ is defined by: 
\begin{eqnarray}\label{q-rel-entr}
S(\psi,\vphi):=\left\{
  \begin{array}{ll}
\textrm{tr}_{{\cal A}}D_{\psi}
(\log D_{\psi}- \log D_{\vphi}),&
 \textrm{if supp} 
(\psi) \leq \textrm{supp}(\vphi)
\\ 
\infty,&  \textrm{otherwise.}
  \end{array}
\right. 
\end{eqnarray}
Observe that in the case of a commutative $C^*$-algebra ${\cal A}$ the
quantum relative entropy $S$ coincides with the classical relative
entropy $H$ defined in (\ref{rel-entr}), where the probabilities are defined\ as the
expectations of minimal projectors in $\mathcal{A}$ . The functional $S(\cdot, \vphi)$ is continuous on ${\cal S}({\cal A})$ only if the reference state $\vphi$ is faithful, i.e. $\textrm{supp} \vphi = \idn_{\cal A}$, otherwise it is lower semi-continuous.
The relative entropy distance from the
reference state $\vphi$ to a subset $\Psi \subseteq {\cal S}({\cal
  A})$ is given by:
\begin{eqnarray}\label{q-rel-entr-dist}
S( \Psi,\vphi):= \inf_{\psi \in \Psi}S(\psi, \vphi). 
\end{eqnarray}
Now we are in the position to state our main result:
\begin{theorem}[Quantum Sanov Theorem]\label{q-sanov}
Let $\vphi \in {\cal S}({\cal A})$ and $\Psi \subseteq {\cal S}({\cal A})$. 
There exists a sequence $\{p_{n}\}_{n \in \nn}$ of orthogonal projections $p_n \in {\cal A}^{(n)}$ such that
\begin{eqnarray}\label{condition}
\lim_{n \to \infty} \psi ^{\otimes n}(p_n)=1, \qquad \forall \psi \in \Psi,
\end{eqnarray}
and
\begin{eqnarray*} 
\lim_{n\to \infty} \frac{1}{n} \log \vphi^{\otimes n}(p_n)= - \inf_{\psi \in \Psi} S(\psi, \vphi). 
\end{eqnarray*}
Moreover, for each sequence of projections $\{\tilde p_n\}$ fulfilling
(\ref{condition}) we have
\begin{eqnarray*}
  \liminf_{n\to \infty}\frac{1}{n}\log \vphi^{\otimes n}(\tilde
  p_n)\ge - \inf_{\psi \in \Psi} S(\psi, \vphi),
\end{eqnarray*}
such that $S(\Psi,\varphi)$ is the best achievable separation rate.
\end{theorem}
The proof of Theorem \ref{q-sanov} will be based to a large extent on the
following classical lemma, which is a stronger version of Theorem \ref{class-result}. 
\begin{lemma}\label{class-result2}
Let $Q \in {\cal P}(A)$ and $\Omega \subseteq {\cal P}(A)$. For each
sequence $\{\eps_n\}_{n\in\nn}$ satisfying $\eps_{n}\searrow 0$ and $\frac{\log(n+1)}{n\eps_{n}^{2}}\to 0 $ there exists a sequence $\{M_n\}_{n \in
  \nn}$ of subsets $M_n \in A^n$ such that for each $P\in \Omega$
there is an $N(P)\in \nn$ with
\begin{eqnarray}\label{eq:prob1}
P^n(M_n)\ge 1- (n+1)^{\# A}\cdot 2^{-nb\eps_{n}^{2}},\qquad \forall n\ge N(P),
\end{eqnarray}
where $b$ is a positive number. Moreover we have:
\begin{enumerate}
\item
$\liminf_{n \to \infty}\frac{1}{n} \log Q^n (M_n)\geq - H(\Omega, Q)$, 
\item
$ Q^n (M_n) \leq (n+1)^{\#A}\cdot 2^{- n I_n}, \qquad \forall n \in
\nn$, \\ where $I_n \geq 0$ and for all $n\in \nn $ fulfilling $\eps_n\le\frac{1}{2}$
\begin{eqnarray}\label{eq:fin-In}
 0\le H(\Omega, Q)- I_n\le \log(\# A)\eps_n -\eps_n\log \eps_n -\eps_n \log Q_{min},
\end{eqnarray}
holds with  $ Q_{min}:= \min \{Q(a): \ Q(a)>0,\ a \in A\}$.
\end{enumerate}
\end{lemma}
\textbf{Proof:} 
Due to the classical Stein's lemma any sequence of subsets $\{M_n\}_{n \in \nn}$, 
which has asymptotically a non vanishing measure with respect to the product distributions $P^{n}$ satisfies:
\begin{eqnarray}\label{Stein-lower-bound}
\liminf_{n \to \infty} \frac{1}{n} Q^n (M_n) \geq -H(P, Q),
\end{eqnarray} 
where $Q \in {\cal P}(A)$ is the reference distribution. 
Then the lower bound (\ref{Stein-lower-bound}) implies the first item  of
lemma (\ref{class-result2}).

We partition the set $\Omega$ into the set $\Omega_1$ consisting of 
probability distributions which are absolutely continuous w.r.t. 
$Q$ and its complement $\Omega_2$ within $\Omega$, i.e.
\begin{displaymath}
  \Omega_1:=\{P\in \Omega: H(P,Q)<\infty\},\quad \textrm{ and }\quad \Omega_2:=\Omega_{1}^{c}\cap \Omega.
\end{displaymath}
 Observe that
\begin{eqnarray}\label{eq:infi}
H(\Omega,Q) = \left\{
  \begin{array}{ll}
H(\Omega_1,Q),&
 \textrm{if } 
\Omega_1 \neq \emptyset
\\ 
\infty,&  \textrm{otherwise.}
  \end{array}
\right. 
\end{eqnarray} 
holds.  We will treat these two sets separately.

It is obvious that we can ideally distinguish the distributions in $\Omega_2$ from $Q$, we just have to set
\begin{eqnarray}\label{eq:m2}
 M_{2,n}:=\{x^{n}\in A^n: Q^{n}(x^{n})=0\textrm{ and } P^{n}(x^{n})>0 \textrm{ for some }P\in \Omega_2\}. 
\end{eqnarray}
Then we have for all $n\in \nn$
\begin{eqnarray}\label{eq:pm2}
 Q^{n}( M_{2,n})=0.  
\end{eqnarray}
Moreover we have for each $P\in \Omega_2 $ and $n\in \nn$
\begin{eqnarray}\label{eq:qm2}
P^{n}(M_{2,n})= 1-q_{P}^{n}\to 1\quad (n\to \infty),
\end{eqnarray}
where
\begin{displaymath}
q_{P}:=P(A_+),
\end{displaymath}
with 
\begin{eqnarray}\label{Aplus}
A_+:=\{a  \in A: Q(a) > 0\}.
\end{eqnarray}
Observe that the speed of convergence in (\ref{eq:qm2}) is \emph{exponential}. 

In treating the set $\Omega_1$ we may consider the restricted alphabet
$A_+$ defined in (\ref{Aplus}) only. 
Note that $H(\cdot ,Q)$ is continuous as a functional on ${\cal P}(A_+)$.  Choose a
sequence $\eps_{n}\searrow 0$ with $\frac{\log(n+1)}{n\eps_{n}^{2}}\to 0 $ and define the
following decreasing family of sets

\begin{displaymath}
  \Omega_{n}:=\{R\in {\cal P}(A_+):\Vert R-P\Vert_1 \leq \eps_{n}  \textrm{ for at least one }P\in \Omega_1\}.
\end{displaymath}
Observe that $ \Omega_{n}\searrow \overline{\Omega_{1}} $. Moreover we set
\begin{eqnarray}\label{eq:Mn}
M_{1,n}:=\{ x^n\in A_+^n:P_{x^n}\in \Omega_n\},
\end{eqnarray}
where $P_{x^n}$ denotes the empirical distribution or type of the sequence
$x^n$. Now, by type counting methods (cf. \cite{Thomas_Cover} section
12.1) and Pinsker's inequality
$H(P_{1},P_{2})\ge\frac{1}{2\ln 2}\Arrowvert
P_{1}-P_{2}\Arrowvert_{1}^{2}$ we arrive at
\begin{eqnarray}\label{eq:proof-prob1}
P^{n}(M_{1,n}^{c})\le (n+1)^{\# A_+}2^{-nb\eps_{n}^{2}}\to 0\quad (n\to \infty),
\end{eqnarray}
for each $P\in \Omega _{1}$ where $b$ is a positive number and $M_{n}^{c} $ denotes the complement
of the set $M_{n} $.\newline
The upper bounds with respect to the distribution $Q$ are a consequence of type counting methods together with
the (lower semi-) continuity of the functional $H( \cdot , Q)$ combined
with (\ref{eq:infi}) and the fact that $\Omega_{n}\searrow \overline{\Omega_{1}} $:
\begin{eqnarray}\label{eq:firstupper}
Q^n (M_{1,n}) \leq (n+1)^{\# A_+}2^{-n H(\Omega_n, Q)}
\quad\textrm{(by type counting, cf. \cite{Thomas_Cover} sect. 12.1)}.  
\end{eqnarray}
We set $I_{n}:= H(\Omega_n, Q) $. Observe that the
sequence $(I_{n})_{n\in\nn}$ is increasing and $ I_n \le \min_ {P\in
  \overline{\Omega_{1}}} H(P, Q) $, for all $n\in\nn$, since $ \Omega_{n}\searrow \overline{\Omega_{1}} $. 
Next, observe that
\begin{eqnarray}\label{eq:closure}
\overline{\Omega_{n}}\subseteq \{R\in {\cal P}(A_+):\Arrowvert R-P\Arrowvert_{1}\leq \eps_{n} \quad
  \textrm{for at least one }P\in \overline\Omega_{1} \}.
\end{eqnarray}
Let $R_n\in \overline{\Omega_{n}}$ be such that 
$H(R_n ,Q)=\min_{R \in \overline \Omega_n}H(R, Q)$. By the continuity we have $H(R_n, Q) = I_n$. According to (\ref{eq:closure}) for each $n\in\nn$
there is a distribution $P_{n}\in\overline{\Omega_{1}}$ such that $\Arrowvert R_n
-P_n\Arrowvert_{1}\le \eps_{n}$. Using the inequality $|H(P)-H(R)|\le
\log(\# A)\Arrowvert P-R\Arrowvert_{1}+\eta ( \Arrowvert
P-R\Arrowvert_{1})$ valid for distributions $P,R$ with $\Arrowvert
P-R\Arrowvert_{1}\le \frac{1}{2}$, where $\eta(t):=-t\log t$, and $Q_{min}:=\min \{Q(a): a \in A_+\} $, we obtain finally
\begin{eqnarray*}
 0\le H(\Omega,Q)-I_n  &=& H(\Omega_{1},Q)-I_n\ \  \quad\quad\quad \textrm{(by
   (\ref{eq:infi}))}\nonumber \\ &=& \min_{Q \in \overline \Omega_{1}}H(P, Q)- I_n \qquad \textrm{(by continuity)} \nonumber \\ & \le & H(P_n,Q)-H(R_n,Q)\nonumber \\ &=& H(R_n) - H(P_n) + \sum_{a \in A_+} (R_n(a)- P_n(a))\log P(a)\nonumber \\
& \le & \log(\# A_+)\Arrowvert P_n-R_n\Arrowvert_{1}+\eta ( \Arrowvert
P_n-R_n\Arrowvert_{1})\nonumber \\
& & - \Arrowvert P_n-R_n\Arrowvert_{1}\log Q_{min}\nonumber \\
& \le & \log(\# A)\eps_n -\eps_n\log \eps_n - \eps_n \log Q_{min}.\nonumber
\end{eqnarray*}
Now,
setting
\begin{displaymath}
  M_n:= M_{1,n}\cup M_{2,n},
\end{displaymath}
we see by (\ref{eq:pm2}) and (\ref{eq:firstupper}) that for all
$n\in\nn$ we have
\begin{displaymath}
Q^{n}( M_n)\le Q^{n}( M_{1,n})+ Q^{n}( M_{2,n})\le (n+1)^{\# A}2^{-nI_{n}}.
\end{displaymath}
Moreover for each $P\in \Omega$ we may infer from (\ref{eq:qm2}) and (\ref{eq:proof-prob1}) that
for all sufficiently large $n\in\nn$
\begin{displaymath}
  P^{n}(M_n)\ge 1- (n+1)^{\# A}2^{-nb\eps_{n}^{2}},
\end{displaymath}
holds.$\qquad\Box$


\section{Proof of the quantum Sanov theorem}
Before we prove the quantum Sanov theorem \ref{q-sanov} we cite the here relevant known results.
We define the maximal separating exponent
\begin{eqnarray*}
\beta_{\eps, n}(\psi^{\otimes n}, \vphi^{\otimes n}):= \min \{\log \vphi^{\otimes n}(q): q \in {\cal A}^{(n)} \textrm{ projection},\ \psi^{\otimes n}(q) \geq 1-\eps   \}.
\end{eqnarray*}
\begin{proposition}\label{proposition}
Let $\psi, \vphi \in {\cal S}({\cal A})$ with the relative entropy $S(\psi, \vphi)$. Then for every $\eps \in (0,1)$
\begin{eqnarray}
\lim_{n \to \infty}\frac{1}{n} \beta_{\eps,n}(\psi^{\otimes n}, \vphi^{\otimes n})= -S(\psi, \vphi).
\end{eqnarray} 
\end{proposition}
The assertion of Proposition \ref{proposition} was shown by Ogawa and Nagaoka in \cite{ogawa}. A different proof based on the approach of Hiai and Petz in \cite{petz} was given in \cite{bs2}. 
\\
\textbf{Proof of the main Theorem \ref{q-sanov}:}\\
\textsl{1. Proof of the lower bound:}
Due to the Proposition \ref{proposition} any sequence of projections $\{p_n\}_{n \in \nn}$, which has asymptotically a non vanishing expectation value with respect to the stationary product state $\{\psi^{\otimes n}\}_{n \in \nn}$ satisfies:
\begin{eqnarray}\label{ind-lower-bound}
\liminf_{n \to \infty} \frac{1}{n}\log \vphi^{\otimes n}(p_n)\geq -S(\psi, \vphi),
\end{eqnarray}
where $\vphi \in {\cal S}({\cal A})$ is a fixed reference state.\\
The lower bound (\ref{ind-lower-bound}) implies the lower bound 
\begin{eqnarray*}
\liminf _{n \to \infty} \frac{1}{n} \log \vphi^{\otimes n}(p_n) \geq - S(\Psi, \vphi)
\end{eqnarray*}
for any sequence $\{p_n\}_{n \in \nn}$ of orthogonal projections $p_n \in {\cal A}^{(n)}$ satisfying condition  (\ref{condition}) in Theorem \ref{q-sanov}. 
\\ \\
\textsl{Proof of the upper bound:}
To obtain the upper bound
\begin{eqnarray*}
\limsup_{n \to \infty} \frac{1}{n}\log \vphi^{\otimes n}(p_n)\leq -S(\Psi, \vphi),
\end{eqnarray*}
where $\vphi$ is a fixed reference state, it is obviously sufficient to show that to each positive $\delta$ there exists a sequence $p_n$ such that
\begin{eqnarray}
\lim_{n \to \infty} \psi ^{\otimes n}(p_n)=1, \qquad \forall \psi \in \Psi,
\end{eqnarray}
and
\begin{eqnarray*}
\limsup_{n \to \infty} \frac{1}{n}\log \vphi^{\otimes n}(p_n)\leq -S(\Psi, \vphi)+\delta
\end{eqnarray*}
is fulfilled for sufficiently large $n$.
To show this we will apply the classical result, 
Lemma \ref{class-result2}, to states restricted to appropriate abelian 
subalgbras approximating the quasi-local algebra ${\cal A}^{\infty}$. \\
Consider the spectral decomposition of the density operator $D_{\vphi}$:
\begin{eqnarray*}
D_{\vphi}=\sum_{i=1}^d \lambda_i e_i, 
\end{eqnarray*}
where $\lambda_i$ are the eigen-values and $e_i$ are the corresponding
spectral projections.
It follows a decomposition for $D_{\vphi^{\otimes l}}=D_{\vphi}^{\otimes l}$:
\begin{eqnarray*}
D_{\vphi}^{\otimes l} = \sum_{i_1,\dots, i_l=1 }^{d} \left ( \prod_{j=1}^{l}\lambda_{i_j} \right ) \bigotimes _{j=1}^l e_{i_j},
\end{eqnarray*}
which leads to the spectral representation
\begin{eqnarray*}
D_{\vphi}^{\otimes l} = \sum_{l_1, \dots, l_d:\ \sum_i l_i =l} \left ( \prod_{i=1}^{d}\lambda_i^{l_i} \right ) e_{l_1, \dots, l_d},
\end{eqnarray*} 
with the spectral projections
\begin{eqnarray*}
 e_{l_1, \dots, l_d} := \sum_{(i_1, \dots, i_l) \in I_{l_1 \dots l_d}}\bigotimes_{j=1}^l e_{i_j},
\end{eqnarray*}
where $I_{l_1 \dots l_d} := \{(i_1, \dots, i_l): \# \{j: i_j=k\}=l_k \textrm{ for } k\in [1,d]\}.$\\
Let $\psi$ be a state on ${\cal A}$ and $l \in\nn$. We denote by 
$ {\cal D}_{l, \psi}$ the abelian subalgebra of ${\cal A}^{(l)}$ generated by $\{e_{l_1 \dots l_d}\}_{l_1 \dots l_d}\cup \{e_{l_1 \dots l_d}D_{\psi}^{\otimes l} e_{l_1 \dots l_d} \}_{l_1 \dots l_d}$. As a finite-dimensional abelian algebra, it has a representation 
\begin{eqnarray*}
{\cal D}_{l, \psi}= \bigoplus_{i=1}^{d_l} \cc \cdot f_{l,i},
\end{eqnarray*}
where $\{f_{l,i}\}_{i=1}^{d_l}$ is a set of  mutually orthogonal minimal projections in ${\cal D}_{l, \psi}$.\\  
Hiai and Petz have shown that
\begin{eqnarray}\label{hiai-petz}
S(\psi^{\otimes l}, \vphi^{\otimes l}) = S(\psi^{\otimes l} 
\upharpoonright {\cal D}_{l, \psi}, \vphi^{\otimes l} \upharpoonright {\cal D}_{l,\psi}) + S(\psi ^{\otimes l} \circ E_l) - S(\psi^{\otimes l}),
\end{eqnarray}
where $\psi
\upharpoonright {\cal D}$ denotes the restriction of a state $\psi \in {\cal S}({\cal A})$ to a subalgebra ${\cal D} \subseteq {\cal A} $  and $E_l$ is the conditional expectation with respect to the canonical trace in ${\cal A}^{(l)}$:


\begin{eqnarray*}
&E_l :&  {\cal A}^{(l)} \longrightarrow  \bigoplus_{l_1 \dots l_d: \sum_i l_i=l}e_{l_1 \dots l_d} {\cal A}^{(l)} e_{l_1 \dots l_d},  \\ & E_l(a)& := \sum _{l_1 \dots l_d: \sum_i l_i =l} e_{l_1 \dots l_d} a e_{l_1 \dots l_d}.
\end{eqnarray*}
Observe that 
\begin{eqnarray}\label{estimate}
S(\psi ^{\otimes l} \circ E_l) - S(\psi^{\otimes l}) \leq d \log (l+1),\quad \textrm{(cf. \cite{petz}, \cite{bs2})}
\end{eqnarray}
which 
gives the lower bound 
\begin{eqnarray}\label{hiai-petz-2}
S(\psi^{\otimes l} \upharpoonright {\cal D}_{l,\psi}, \vphi^{\otimes l} \upharpoonright {\cal D}_{l, \psi})\geq S(\psi^{\otimes l}, \vphi^{\otimes l}) - d\log (l+1)
\end{eqnarray}
implying 
\begin{eqnarray*}
\lim_{l \to \infty} \frac{1}{l} S(\psi ^{\otimes l} \upharpoonright  {\cal D}_{l, \psi}, \vphi ^{\otimes l}\upharpoonright {\cal D}_{l, \psi}) = S(\psi, \vphi).
\end{eqnarray*}
Next we consider a maximal
abelian refinement 
${\cal B}_{l, \psi}$ of ${\cal D}_{l, \psi}$ in the sense of the algebra ${\cal A}^{(l)}$:
\begin{eqnarray*}
{\cal B}_{l, \psi}:= \bigoplus_{j,k=1}^{d_l, a_{l,j}} \cc \cdot g_{l,j,k},
\end{eqnarray*}
where $g_{l,j,k}$ are one-dimensional projections in the algebra ${\cal A}^{(l)}$ such that $f_{l,j}=\bigoplus_{k=1}^{a_{l,j}} \cc \cdot g_{l,j,k}$. This means that ${\cal B}_{l, \psi} \supseteq {\cal D}_{l, \psi}$.
 It holds by monotonicity of the relative entropy and by the estimate (\ref{hiai-petz-2})
\begin{eqnarray}\label{lower-bound}
S(\psi^{\otimes l} \upharpoonright {\cal B}_{l,\psi}, \vphi^{\otimes l} \upharpoonright {\cal B}_{l, \psi}) &\geq& S(\psi^{\otimes l} \upharpoonright {\cal D}_{l,\psi}, \vphi^{\otimes l} \upharpoonright {\cal D}_{l, \psi}) \nonumber \\ &\geq& l(S(\psi, \vphi) - \eta_l), 
\end{eqnarray}
where we used the abbreviation $\eta_{l}:=   \frac{d \log (l+1)}{l}$ in the last line.

Due to the Gelfand isomorphism and the Riesz representation theorem the restricted  states $\psi^{\otimes l}\upharpoonright {\cal B}_{l, \psi} $ and $\vphi^{\otimes l}\upharpoonright {\cal B}_{l, \psi} $ can be identified with probability measures $P$ and $Q$ on the compact maximal ideal space $B_{l, \psi}$ corresponding to the $d^l$-dimensional abelian algebra ${\cal B}_{l, \psi}$. The relative entropy of $P$ with respect to $Q$ is determined by:
\begin{eqnarray*}
H(P,Q)=S(\psi^{\otimes l}\upharpoonright {\cal B}_{l, \psi},\vphi^{\otimes l}\upharpoonright {\cal B}_{l, \psi})\geq l\cdot (S(\psi, \vphi)-\eta_{l}).
\end{eqnarray*}
Similarly, the states $\psi^{\otimes nl}\upharpoonright {\cal B}_{l, \psi}^{(n)}$ and $\vphi^{\otimes nl}\upharpoonright {\cal B}_{l, \psi}^{(n)}$ correspond to the product measures $P^n$ and $Q^n$ on the product space $B_{l, \psi}^n$.
\\ We define
\begin{eqnarray*}
S_l:=\inf \{ S(\psi^{\otimes l}\upharpoonright {\cal B}_{l,\psi}, \vphi^{\otimes l} \upharpoonright {\cal B}_{l,\psi}): \psi \in \Psi  \}
\end{eqnarray*}
and fix an $\psi_0 \in \Psi$. 
For any $\psi \in \Psi$ and each $l\in\nn$
there exists a unitary operator $U_{\psi} \in {\cal A}^{(l)}$ that transforms
the minimal projections spanning ${\cal B}_{l, \psi}$ into the
minimal projections of ${\cal B}_{l, \psi_{0}}$ and that leaves the
spectral subspaces of $D_{\vphi^{\otimes l}}$ invariant. Let us denote
by $\mathfrak{U}_{l}(\Psi,\vphi)$ the set of unitaries having these
properties. To each $\psi \in \Psi$ denote by $\tilde \psi^{(l)}$ the state on ${\cal A}^{(l)}$ with density operator $U_{\psi}D_{\psi}^{\otimes l}U^*_{\psi}$. Then we have
\begin{eqnarray*}
S(\psi^{\otimes l}\upharpoonright {\cal B}_{l,\psi}, \vphi^{\otimes l} \upharpoonright {\cal B}_{l,\psi})=S(\tilde \psi^{(l)}\upharpoonright {\cal B}_{l,\psi_0}, \vphi^{\otimes l} \upharpoonright {\cal B}_{l,\psi_0}).
\end{eqnarray*}   
Let $\Omega_{l}$ be the set of probability measures on $B_{l, \psi_0}$ corresponding to all $\tilde \psi^{(l)}\upharpoonright {\cal B}_{l,\psi_0}$, where $\psi \in \Psi$. Further let the measure $Q$ on $B_{l, \psi_0}$ correspond to the restricted reference state $\vphi^{\otimes l} \upharpoonright {\cal B}_{l,\psi_0}$.
Then
\begin{eqnarray}\label{lower-bound-h}
H(\Omega_{l}, Q) \geq S_l \geq l \cdot (S(\Psi, \vphi) - \eta_l),
\end{eqnarray}
where the second inequality follows from (\ref{lower-bound}).
Due to the Lemma \ref{class-result2} there exists a sequence $\{M_n\}_{n \in \nn}$ of subsets $M_n \in B_{l,\psi_0}^{n}$ (cf. (\ref{eq:Mn}))  such that
\begin{eqnarray}\label{class1}
\lim_{n \to \infty } P^{n}(M_n) = 1, \qquad \forall P \in \Omega_{l}
\end{eqnarray}
and for every $n \in \nn$
\begin{eqnarray*}
 Q^{n}(M_n) \leq (n+1)^{d^{l}} 2^{-n I_n(l)},
\end{eqnarray*}
where $I_n(l) \nearrow H(\Omega_l, Q)$ for $n \to \infty$. Moreover, we know that  $I_n(l)\geq H(\Omega_l, Q) - \eps_n (\log d^l - \log \eps_n - \log Q_{\min}(l) )$, where $Q_{\min}(l):=\min\{Q(a): a \in B_{l, \psi_0}\}$. We introduce the abbreviation $\vartriangle _n ^{(l)}:= \eps_n (\log d^l - \log \eps_n - \log Q_{\min}(l) )$. It holds: 
\begin{eqnarray}\label{class2}
\frac{1}{n} \log Q^{n}(M_n) \leq \frac{d^l \log (n+1)}{n} - I_n \leq  \frac{d^l \log (n+1)}{n} - (H(\Omega_l,Q)- \vartriangle _n ^{(l)}).
\end{eqnarray} 
To each $M_n \in B_{l,\psi_0}^n$ there corresponds a projection $p_{ln}$ in $ {\cal B}_{l, \psi_0}^{\otimes n} \subseteq {\cal A}^{(nl)}$. For an arbitrary $m \in \nn$ such that $m=nl+r \in \nn$ with $r\in \{0, \dots, l-1\}$ we define a projection $p_{m} \in {\cal A}^{(m)}$ by
\begin{eqnarray*}
p_m:=p_{nl} \otimes \idn_{[nl+1,nl+r]},
\end{eqnarray*}
where $\idn_{[nl+1,nl+r]}$ denotes the indentity in the local algebra ${\cal A}_{[nl+1,nl+r]}$. It holds 
\begin{eqnarray*}
 \tilde \psi^{(l)\otimes n}(p_{nl})=P^{n}(M_n), \qquad \forall \psi \in  \Psi 
\end{eqnarray*}
and
\begin{eqnarray*}
\frac{1}{m}\log \vphi^{\otimes m}(p_m)\leq \frac{1}{nl}\log \vphi^{\otimes nl}(p_{nl})= \frac{1}{nl}\log Q^{n}(M_n). 
\end{eqnarray*}
Using (\ref{class1}), (\ref{class2}) and (\ref{lower-bound-h}) we conclude
\begin{eqnarray}\label{bed1}
\lim_{n \to \infty} \psi^{\otimes nl}(U_{\psi}^{* \otimes n}p_{nl}U_{\psi}^{\otimes n})=1, \qquad \forall \psi \in \Psi 
\end{eqnarray}
and 
\begin{eqnarray}
\frac{1}{m}\log \vphi^{\otimes m}(p_m)&=& \frac{1}{nl}\log Q^n(M_n) \nonumber \\ &\leq& \frac{d^l
  \log(n+1)}{nl} - \frac{1}{l}I_n(l) \nonumber \\ &\leq& \frac{d^l
  \log(n+1)}{nl} - \frac{1}{l} (H(\Omega_l,Q) - \vartriangle_n^{(l)})  \nonumber \\ &\leq&
\frac{d^l \log(n+1)}{nl} - S(\Psi, \vphi)+\eta_{l} + \frac{\vartriangle_n^{(l)}}{l} \label{bed2}.
\end{eqnarray}
For fixed $l\in\nn$ we construct for each $n  \in \nn$ the projection:
\begin{eqnarray*}
\overline p_{nl} := \bigvee_{U \in \mathfrak{U}_{l}(\Psi,\vphi)} U^{* \otimes n} p_{nl} U^{ \otimes n}.
\end{eqnarray*}
For an arbitrary number $m=nl+r$, $r \in \{0,\dots,l-1\}$, we define
\begin{eqnarray*}
\overline p_m:= \overline p_{nl} \otimes \idn_{[nl+1,nl+r]}.
\end{eqnarray*}
It follows for arbitrary $\psi \in \Psi$ and each $m=nl+r \in \nn$:
\begin{eqnarray}\label{psi-estimate}
\psi^{\otimes m}(\overline p_m)= \psi^{\otimes nl}(\overline p_{nl})\geq \psi^{\otimes nl}(U_{\psi}^{* \otimes n}p_{nl} U_{\psi}^{\otimes n}).
\end{eqnarray}
Using the estimate (\ref{bed1}) we obtain the general statement:
\begin{eqnarray*}
\lim_{m \to \infty} \psi^{\otimes m}(\overline p_{m})=1, \qquad \forall \psi \in \Psi.
\end{eqnarray*}
Next we consider the expectation values $\vphi^{\otimes nl}(U^{* \otimes n}p_{nl}U^{\otimes n})$ for any $U \in \mathfrak{U}_{l}(\Psi,\vphi)$ and $n \in \nn$. From the assumed invariance of $D_{\varphi^{\otimes l}}$ with respect to the unitary transformations given by elements of $\mathfrak{U}_{l}(\Psi,\vphi)$ we conclude
\begin{eqnarray}\label{u-inv-expect}
\vphi^{\otimes nl}(U^{* \otimes n}p_{nl}U^{\otimes n})=\vphi^{\otimes nl}(p_{nl}), \qquad \forall U \in \mathfrak{U}_{l}(\Psi,\vphi).
\end{eqnarray}
The dimension of the symmetric subspace 
\begin{eqnarray*}
\textrm{SYM}({\cal A}^{(l)},n):= \textrm{span}\{A^{\otimes n}: A \in {\cal A}^{(l)}\}
\end{eqnarray*}
is upper bounded by $(n+1)^{\dim{\cal A}^{(l)}}$, which leads to the  estimate
\begin{eqnarray}\label{sym-ur}
\textrm{tr }\overline p_{nl}\leq  (n+1)^{d^{2l}} \cdot \textrm{tr } p_{nl}.
\end{eqnarray}
Using (\ref{u-inv-expect}), (\ref{sym-ur}) and (\ref{bed2}) we obtain
\begin{eqnarray}\label{eq:final}
\frac{1}{m}\log \vphi^{\otimes m}(\overline p_{m}) &\leq&
\frac{1}{nl}\log \vphi^{\otimes nl}(\overline p_{nl}) \nonumber \\ 
&\leq& \frac{1}{nl} \log ((n+1)^{d^{2l}}\cdot \vphi^{\otimes nl}(
p_{nl} ))\nonumber \\
&\leq& \frac{(d^{2l}+d^{l}) \log (n+1)}{nl}-S(\Psi, \vphi)+\eta_{l} + \frac{\vartriangle_n^{(l)}}{l}.
\end{eqnarray}
For fixed $l$ the upper bound above converges to $-S(\Psi, \vphi) + \eta_l$, for $n \to \infty$.
Choosing $l$ sufficiently large, $\eta_l$ becomes smaller than
$\delta$. This proves the upper bound.
$\qquad \Box$
\section{Two examples}
1. Consider a quantum system where $\mathbb{C}^{2}$ is the underlying Hilbert space
and let $v,w$ be two different non-orthogonal unit vectors in $%
\mathbb{C}^{2}$. Let $\psi^{\otimes n} $ be the product state on $(\frak{B}(\mathbb{C}%
^{2}))^{\otimes n}$ with the density operator $p_{w}^{\otimes n}$, where $p_w$ is the projection onto the one-dimensional subspace in $\mathbb{C}^2$ spanned by $w$. Further
let $\delta \geq 0$ and denote by $\vphi _{\delta }$ the state on $
\frak{B}(\mathbb{C}^{2})$ corresponding to the density operator $(1-\delta )p_{v}+\delta p_{w}$. It
seems rather clear that any reasonable attempt to define empirical
distributions (states) in quantum context should choose $p_{w}$
in the case of $\psi^{\otimes n} $ (or more general the underlying one-site state in the
case of a stationary product state). So, when trying to define typical
projectors via empirical states and to use these in analogy to the classical
Sanov's theorem, the $n$-block typical projector $p^{(n)}$ for the set $%
\Psi=\{\psi \} \subseteq {\cal S}(\frak{B}(\mathbb{C}%
^{2}))$ would be expected to fulfil $p^{(n)}\geq p_{w}^{\otimes n}$. Then we have $\vphi _{\delta }^{\otimes n}(p^{(n)})\geq
\vphi _{\delta }^{\otimes n}(p_{w}^{\otimes n})=(\delta +(1-\delta
)\langle v,w\rangle ^{2})^{n}\geq \langle v,w\rangle ^{2n}$. On the other
hand, the relative entropy of the density operator $p_{w}$ with repect to $%
(1-\delta )p_{v}+\delta p_{w}$ can be made arbitrary
large by choosing $\delta $ small but positive. This shows that, in contrast
to the classical situation, when relying on empirical states the relative
entropy rate is \emph{not} an accessible separation rate (which can be at
most $-2\log |\langle v,w\rangle | $) . We might simplify the argument by
saying that though the relative entropy of $\psi$ with respect to $%
\vphi $ is infinite the separation rate using empirical distributions
remains bounded. But choosing $p^{(n)}$ as $p_{(v^{\otimes
n})^{\perp }}$, where $(v^{\otimes n})^{\perp }$ denotes the orthogonal
complement of the vector $v^{\otimes n}$ in $(\mathbb{C}^{2})^{\otimes n}$ yields $\vphi
^{\otimes n}(p^{(n)})\equiv 0$ and $\psi ^{\otimes n}(p^{(n)})\rightarrow 1$, hence
the separation rate can in fact be made infinite when choosing the typical
projector in another way.

2. A slightly more involved example shows that, again in contrast to the
classical case, there is in general no \emph{universal }choice of the
separating projector, i.e. it has to depend upon the reference state $\vphi $%
. This time we will refer directly to the infinite relative entropy case and
leave the simple 'smoothening' argument which leads to a finite entropy
example to the reader. Let $v$ and $w$ be two orthogonal unit vectors in $\mathbb{%
C}^{2}$. Let $\Phi$ be the set of pure states $\vphi _{t}$ on $
\frak{B}(\mathbb{C}^{2})$ corresponding to the vectors $v_{t}:=\cos t\cdot v+\sin t\cdot w,t\in
\lbrack -T,T],\frac{\pi }{2}>T>0$ and $\Psi=\{\psi \}$, where $%
\psi $ is the pure state corresponding to $w$. Assume there is a typical
projector $p^{(n)}$ for $\psi ^{\otimes n}$ separating it from each $\vphi
_{t}^{\otimes n}$ super-exponentially fast. This should be valid for an universal
projector since all the relative entropies $S(\psi,\vphi
_{t})$ are infinite. Let SYM$(n)\subset (\mathbb{C}^{2})^{\otimes n}$
be the symmetrical $n$-fold tensor product of  $\mathbb{C}^{2}$. Without any
loss of the generality we may choose $p^{(n)} \leq p_{\textrm{SYM%
}(n)}$ since all $v_{t}^{\otimes n}$ as well as $w^{\otimes n}$ belong to SYM%
$(n)$. Observe that the existence of $p^{(n)}$ (with the desired property)
implies the existence of at least one (sequence of) unit vectors $x_{n}$ in
SYM$(n)$ such that $\langle x_{n},v_{t}^{\otimes n}\rangle $ tends to zero
super-exponentially fast uniformly in $t$. 
Choose an orthonormal basis in SYM$(n)$ by $e_{n,k}:=\binom{n}{k}%
^{1/2}/n!\sum_{\pi \in \textrm{PERM}(n)}U_{\pi }(w^{\otimes k}\otimes
v^{\otimes (n-k)})$, where PERM$(n)$ is the group of $n$-Permutations and $%
U_{\pi }$ is the unitary operator which interchanges the order in the tensor
product according to $\pi $. Representing $v_{t}^{\otimes n}$ in that basis
yields the numerical vector $(\binom{n}{k}^{1/2}(\sin t)^{k}(\cos
t)^{n-k})_{k=0}^{n}$. So the question is whether there exists a sequence of
unit vectors $x_{n}=(x_{n,k})$ such that $\sup_{t\in \lbrack -T,T]}(\cos
t)^{n}\sum_{k}x_{n,k}\binom{n}{k}^{1/2}(\tan t)^{k}$ tends to zero
super-exponentially fast. Observe that the factor $(\cos t)^{n}$ is bounded
from below by $(\cos T)^{n}$ and can be omitted since it goes to zero only
exponentially fast. Moreover, if we replace $x_{n}$ by $\widehat{x}%
_{n}=(x_{n,k}\binom{n}{k}^{-1/2})$ we change its norm only by an at most
exponentially smaller factor (the maximum of binomial coefficient is of
exponential order $2^{n}$). So we may simplify the problem by asking whether
there is a sequence of unit vectors $x_{n}$ which has a super-exponentially
decreasing inner product with the numerical vectors $((\tan t)^{k})_{k=0,1,...,n}$%
, uniformly in $t\in \lbrack -T,T]$. This can be excluded: Let $n$ be uneven
and consider the set of values $t_{m}=\arctan ((1-2\frac{m}{n})\cdot \tan
T),m=0,1,...,n$. Even for this finite set of values we have necessarily $%
\sup_{m}\sum_{k}x_{n,k}(\tan t_{m})^{k}$ $=\sup_{m}\sum_{k}x_{n,k}((1-2\frac{%
m}{n}))^{k}(\tan T)^{k}$ tending to zero at most exponentially fast. In
fact, the factor $(\tan T)^{k}$ can be omitted as before. Let  $V_{n}$ be
the Vandermonde matrix $(((1-2\frac{m}{n}))^{k})_{m,k=0}^{n}$. Then the $%
L_{\infty }$-norm of the vector $V_{n}x_{n}$ can be estimated by a
sub-exponential factor times its $L_{2}$-norm, and by \cite{gautschi}, Example
6.1 the least singular value  of $V_{n}$ behaves like $\pi e^{\frac{\pi }{4}%
}e^{-n(\frac{\pi }{4}+\frac{1}{2}\ln 2)}$.\\

\emph{Acknowledgements.} This work was supported by the DFG via the
project ``Entropie, Geometrie und Kodierung gro\ss er
Quanten-Informationssysteme'', by the ESF via the project ``Belearning'' at the TU Berlin and
the DFG-Forschergruppe ``Stochastische Analysis und gro\ss e
Abweichungen'' at the University of Bielefeld. 

\end{document}